\documentclass[a4paper,11pt]{article}
\usepackage[utf8]{inputenc}
\usepackage{graphicx}
\usepackage{amsmath,amsfonts,amssymb}
\usepackage{hyperref}
\usepackage{color}
\usepackage{cite}
\usepackage{csquotes}
\usepackage{authblk}
\usepackage[toc,page]{appendix}


\title{%
\large\bf Pathologies of the Kimber-Martin-Ryskin prescriptions for unintegrated PDFs: Which prescription should be preferred?}
\author{Benjamin Guiot\footnote{benjamin.guiot@usm.cl}}
\affil[]{\small{Departamento de F\'isica, Universidad T\'ecnica Federico Santa Mar\'ia; Casilla 110-V, Valparaiso, Chile}}
\date{}

\begin{document}

\maketitle

\begin{abstract}
We discuss the different Kimber-Martin-Ryskin (KMR) prescriptions for unintegrated parton distribution functions (uPDFs). We show that the strong-ordering (SO) and the angular-ordering (AO) cutoffs lead to strong discrepancies between the obtained cross sections. While the result obtained with the AO cutoff overestimates the heavy-flavor cross section by about a factor of 3, the SO cutoff gives the correct answer. We also solve the issue of the KMR uPDFs definitions mentioned in \cite{BGS}, and show that, in the case of the AO cutoff, the KMR uPDFs are ill defined.
\end{abstract}

\newpage

\tableofcontents

\section{Introduction}
Understanding transverse-momentum-dependent parton distribution functions has been a topic of increasing theoretical and experimental interest. Compared to the collinear PDFs, they provide additional information on the transverse dynamics of a parton inside the hadron. Depending on the kinematical range, several formalisms exist. The TMD factorization \cite{JiMaYu1,JiMaYu2,BaBoDi,col} is valid for small $k_t/Q$, where $k_t$ is the parton transverse momentum and $Q$ is the hard scale of the process. The TMD PDFs, mainly studied in semi-inclusive deep inelastic scattering and Drell-Yan experiments, provide a 3-dimensional information on the hadron structure and could help to solve the proton-spin crisis. The $k_t$ factorization, first developed in Refs. \cite{ktFac,ktFac2,semihard1,semihard2}, is used at small-x. In this case, $k_t$ is not restricted to small values. It finds applications at the LHC, where the transverse momentum of incoming spacelike partons can indeed be large, due to partonic evolutions.\\

In the context of $k_t$ factorization, where the transverse momentum PDFs are generally refereed as unintegrated PDFs (uPDFs), a popular construction of these functions is given by the Kimber-Martin-Ryskin (KMR) and Watt-Martin-Ryskin (WMR) prescriptions \cite{KMR1,KMR2}. The KMR/WMR uPDFs are usually used with the angular-ordering cutoff (see section \ref{discc} for a more detailed discussion on the different cutoffs), and give a satisfying description of the D mesons $p_t$ distribution, taking into account only the $gg\rightarrow c\bar{c}$ process. 

However, calculations using a variable-flavor-number scheme\footnote{The KMR/WMR uPDFs are generally built from collinear PDFs extracted using the variable-flavor-number scheme.} should by definition include the other processes, in particular the flavor excitation process $Qg\rightarrow Qg$, where $Q$ is a heavy quark. It has been shown that, at leading order, the latter gives the main contribution to the $p_t$ distribution of one heavy quark \cite{gui2}. Consequently, there is necessarily something wrong with calculations that use this scheme, include only the $gg$ contribution, and show a good agreement with data on heavy-quark production. The explanation given in Ref. \cite{gui2} was that these calculations effectively include a large $K$ factor. Naturally, the issue is that after the inclusion of the $Qg$ contribution, the result overestimates the data (an example is shown in Fig. \ref{kmrpt}).\\

In this paper, we present a detailed analysis of the KMR/WMR prescriptions. The angular-ordering WMR uPDFs are used to exemplify the conclusions reached in Ref. \cite{gui2}. We will see that in this case, the effective large $K$ factor is due to the too large $k_t$ tail of the distribution, at $k_t>\mu$, where $\mu$ is the factorization scale. We will discuss several theoretical issues related to the angular-ordering cutoff, and see that the KMR/WMR uPDFs built with the strong-ordering cutoff are free of some of them. Another objective of this paper is to present a solution to the issue of the KMR/WMR uPDFs definitions addressed in Ref. \cite{BGS}. The outline of the paper is as following. After a short review of the Dokshitzer-Gribov-Lipatov-Altarelli-Parisi (DGLAP) equation in section \ref{secdglap}, we present the KMR and WRM prescriptions in section \ref{secKMRPDF}. We will see that they are not equivalent, and that the former does not obey the correct DGLAP equation. In section \ref{discc} we discuss in detail the issue of the KMR/WMR uPDF definitions, related to the fact that apparently mathematically equivalent definitions give different numerical results. Finally, in section \ref{sectrue}, we further study the differences between the KMR/WMR prescriptions, by discussing the angular-ordering (AO) and the strong-ordering (SO) cutoffs. Using different cutoffs leads to significant differences for the cross section, and we will see that the SO cutoff should be preferred. In particular, we show by performing explicit calculations that the SO cutoff gives results compatible with those obtained in Ref. \cite{gui2}.

\section{The Dokshitzer-Gribov-Lipatov-Altarelli-Parisi equation with unregularized splitting functions}\label{secdglap}
In this section, following \cite{ElStWe}, we quickly remind a form of the DGLAP equation that is useful for numerical treatments. For small $\delta x$ and $\delta t$, centered on $x$ and $t$, the variation of the parton density with $t$ is given by
\begin{equation}
\delta f(x,t)=\delta f_{\text{in}}(x,t)-\delta f_{\text{out}}(x,t).\label{dglapun}
\end{equation}
The variable $t$ has the dimension of energy squared. Equation (\ref{dglapun}) simply expresses that the change of a quantity in a volume (here $\delta t \delta x$) is given by what goes in, minus what goes out. Working with one parton flavor, $\delta f_{\text{in}}(x,t)$ receives a contribution from the splitting of partons at $x'>x$:
\begin{eqnarray}
\delta f_{\text{in}}(x,t)&=& \frac{\delta t}{t}\int_x^1 dx'\int_0^1 dz \frac{\alpha_s}{2\pi}\hat{P}(z)f(x',t)\delta(x-zx')\nonumber \\
&=&\frac{\delta t}{t}\int_0^1 \frac{dz}{z} \frac{\alpha_s}{2\pi}\hat{P}(z)f(x/z,t). \label{incont}
\end{eqnarray}
 It is proportional to the parton density at $x'$ multiplied by the probability for a splitting at $t$, with the daughter parton having a fraction $z$ (generally the light-cone momentum fraction) of its mother particle. The delta function ensures that after the splitting, the parton arrives in the volume $\delta t \delta x$. $\hat{P}(z)$ is the unregularized splitting function. Similarly, the outgoing part is given by
\begin{equation}
 \delta f_{\text{out}}(x,t)=\frac{\delta t}{t}f(x,t)\int_0^1 dz \frac{\alpha_s}{2\pi}\hat{P}(z). \label{outcont}
\end{equation}
One of the differences with Eq. (\ref{incont}) is that the parton density is outside of the integral. Indeed, for partons inside the volume $\delta t \delta x$, any splitting will bring them out. So the contribution is simply given by the parton density at $x$ multiplied by the total splitting probability (for fixed $t$).\\

We now consider the realistic case of QCD. The variation of the quark density at leading order reads
\begin{align}\label{delquark}
\delta q(x,t)=&\frac{\delta t}{t}\int_0^1 \frac{dz}{z}\frac{\alpha_s}{2\pi}\left\lbrace \hat{P}_{qq}(z)q\left(\frac{x}{z},t\right)+\hat{P}_{qg}(z)g\left(\frac{x}{z},t\right)\right\rbrace\\ \nonumber
&-\frac{\delta t}{t}q(x,t)\int_0^1 dz \frac{\alpha_s}{2\pi} \hat{P}_{qq}(z).
\end{align}
The case of the gluon density is more complicated. One can arrive in the volume from either $g\rightarrow gg$ or $q\rightarrow gq$, and one leaves the volume from either $g\rightarrow gg$ or $g\rightarrow q\bar{q}$. As explained in Ref. \cite{ElStWe}, one subtlety is that both gluons produced in the splitting $g\rightarrow gg$ can participate, giving
\begin{equation}
\delta g_{\text{in}}(x,t)=\frac{\delta t}{t}\int_0^1 \frac{dz}{z}\frac{\alpha_s}{2\pi}\left\lbrace 2\hat{P}_{gg}(z)g\left(\frac{x}{z},t\right)+\hat{P}_{gq}(z)\left[q\left(\frac{x}{z},t\right)+\bar{q}\left(\frac{x}{z},t\right)\right] \right\rbrace .\label{gin}
\end{equation}
The unregularized splitting functions are given in Ref. \cite{ElStWe}, equations (5.10) and (5.20):
\begin{eqnarray}
 \hat{P}_{gg}(z)&=&C_A\left[\frac{1-z}{z}+\frac{z}{1-z}+z(1-z)\right]\label{pgg} \\
 \hat{P}_{gq}(z)&=&\hat{P}_{gq}(1-z)=C_F\frac{1+(1-z)^2}{z}
\end{eqnarray}
The outgoing part is given by
\begin{equation}
\delta g_{\text{out}}(x,t)=\frac{\delta t}{t}g(x,t)\int_0^1 dz\frac{\alpha_s}{2\pi} \left[ \hat{P}_{gg}(z)+n_f\hat{P}_{qg} \right] . \label{gout}
\end{equation}
Note the factors of $2$ and $1$ in front of $\hat{P}_{gg}$ in Eqs. (\ref{gin}) and (\ref{gout}). The regularized splitting functions are obtained after applying the plus-prescription \cite{ElStWe}:
\begin{equation}
P(z)=\hat{P}(z)_+.
\end{equation}
In the case of the gluon-gluon splitting function the result is
\begin{equation}
P_{gg}(z)=2C_A\left[\frac{z}{(1-z)_+}+\frac{1-z}{z}+z(1-z)\right]+\frac{1}{6}(11C_A-4N_f T_R)\delta(1-z),
\end{equation}
with $T_R=1/2$. Note the factor of 2 in front of $C_A$, compared to the unregularized case. In the following, we will use the unregularized splitting function $\hat{P}_{gg}$, Eq. (\ref{pgg}), with a factor of $2C_A$, for reasons explained in the next section. In the rest of the paper, all of the mentioned splitting functions are unregularized, and they will be written without the "hat", in order to fit with the literature on KMR unintegrated PDFs (uPDFs).

\section{The KMR unintegrated PDFs}\label{secKMRPDF}
We first start by discussing some ambiguities, related to the fact that in the literature, ``KMR formalism" can refer both to Ref. \cite{KMR1} and Ref. \cite{KMR2}. However, the equations given in these papers are \textit{not} equivalent and we will refer to the second one as the  WMR formalism. In Ref. \cite{KMR1}, the DGLAP equation was written as
\begin{equation}
\frac{\partial a(x,\mu^2)}{\partial \ln\mu^2}=\sum_{a'}\frac{\alpha_s}{2\pi}\left[\int_x^{1-\Delta}P_{aa'}(z)a'\left(\frac{x}{z},\mu^2\right)dz-a(x,\mu^2)\int_0^{1-\Delta}P_{a'a}(z)dz \right], \label{kmrdglap}
\end{equation}
where $a(x,\mu^2)=xf_a(x,\mu^2)$ and $f_a(x,\mu^2)$ is the number density. The sum on $a'$ runs over all possible parton flavours: quarks, antiquarks and gluon. Note that in the KMR/WMR prescriptions the gluon-gluon splitting function is defined with a factor of $2C_A$. Comparing Eq. (\ref{kmrdglap}) with Eqs. (\ref{gin}) and (\ref{gout}), we can see that Eq. (\ref{kmrdglap}) does not reproduce the correct DGLAP equation for the gluon density. Indeed, the
coefficients in front of $P_{gg}$ should not be the same. A similar remark also applies for the quark distribution function since, in the last term of Eq. (\ref{kmrdglap}), the sum over $a'$ implies the contribution of both $P_{qq}(z)$ and $P_{gq}(z)$, in disagreement with Eq. (\ref{delquark}).\\

In the WMR case, the DGLAP equation is written with an additional $z$ factor in the last term [see Eq. (17) of Ref. \cite{KMR2}]:
\begin{equation}
\frac{\partial a(x,\mu^2)}{\partial \ln\mu^2}=\sum_{a'}\frac{\alpha_s}{2\pi}\left[\int_x^{1-\Delta}P_{aa'}(z)a'\left(\frac{x}{z},k_t\right)dz-a(x,\mu^2)\int_0^{1-\Delta}zP_{a'a}(z)dz \right]. \label{wmrdglap}
\end{equation}
Consequently, Eqs. (\ref{wmrdglap}) and (\ref{kmrdglap}) are not equivalent. This difference can be traced back to the definition of the Sudakov factor [Eq. (18) in Ref. \cite{KMR2} and Eq. (3) in Ref. \cite{KMR1}]. The extra $z$ factor was justified by saying that it``avoids double-counting the s- and t-channel partons". It was also mentioned that, after integrating over $z$ and summing over $a'$, it gives a factor of $1/2$. In that case, and using $P_{gg}$ with a factor of $2C_A$, Eq. (\ref{wmrdglap}) with $a=g$ is equivalent to Eqs. (\ref{gin}) and (\ref{gout}). It also gives the correct DGLAP equation for the quark, since
\begin{equation}
-\frac{1}{2} q(x,t)\int_0^1 dz \frac{\alpha_s}{2\pi}\left[P_{qq}(z)+P_{gq}(z)\right]=-q(x,t)\int_0^1 dz \frac{\alpha_s}{2\pi}P_{qq}(z).
\end{equation}
Here we used that fact that $P_{qq}$ and $P_{gq}$ are related by $z\rightarrow 1-z$.\\

An advantage of the WMR prescription is that the $z$ factor regularizes the divergence of the splitting function $P_{gg}$ when $z$ goes to zero. In recent papers, the KMR prescription used was in fact the WMR one, as was the case in Ref. \cite{BGS}, which we discuss now.\\

As explained in the introduction, the present work has been motivated by Ref. \cite{BGS}. One of our goals is to discuss the analysis given in that paper. It is then useful to give a short and similar presentation of the WMR formalism, insisting on important details.\\

The goal is to build an unintegrated parton density that obeys (at least approximately)
\begin{equation}
f_a(x,Q^2)=\int_{0}^{Q^2}F_a(x,k_t^2;Q^2)dk_t^2, \label{inttmd}
\end{equation}
with $Q^2$ having the dimension of energy squared\footnote{We chose this notation in order to agree with Ref. \cite{BGS}.}. This equation is sometimes written with a factor of $x$ on the lhs. In this case, the function $F_a(x,k_t^2;Q^2)$ is the momentum density. However, the factor $1/z$ in \cite{BGS}, equation (2), indicates that the authors were working with the parton densities, so we use the relation (\ref{inttmd}). 

The derivation starts with the DGLAP equation. The main trick in the WMR prescription is the observation that by using the Sudakov factor
\begin{equation}
T_a(Q,k_t)=\exp \left\lbrace -\int_{k_t^2}^{Q^2}\frac{dp_t^2}{p_t^2}\sum_{a'}\int_0^{1-\Delta(p_t)}dz\,zP_{a'a}(z,p_t)\right\rbrace, \label{suda}
\end{equation} 
with $P_{a'a}(z,\mu)$ defined by
\begin{equation}
P_{a'a}(z,\mu)=\frac{\alpha_s(\mu^2)}{2\pi}P_{a'a}^{\text{LO}}(z),
\end{equation}
the DGLAP equation\footnote{Strictly speaking, this is not the DGLAP equation since there is an extra $z$ factor in the WMR prescription.} can be rewritten as
\begin{equation}
\frac{\partial}{\partial \ln k_t^2}\left[T_a(Q,k_t)f_a(x,k_t)\right]=T_a(Q,k_t)\sum_{a'}\int_x^{1-\Delta}\frac{dz}{z}P_{aa'}(z,k_t)f_{a'}\left(\frac{x}{z},k_t\right).\label{dglap2}
\end{equation}
However, for this to be correct, one should be careful with the $k_t$ dependence of the Sudakov factor. In particular, as mentioned in Ref. \cite{BGS}, the cutoff $\Delta$ should not be a function of $k_t$ when used in the definition of $T_a$, Eq. (\ref{suda}). In this case, we have
\begin{equation}
\frac{\partial T_a(Q,k_t)}{\partial \ln k_t^2}=T_a(Q,k_t)\sum_{a'}\int_0^{1-\Delta(k_t)}dz\,zP_{a'a}(z,k_t),
\end{equation}
and after a straightforward calculation Eq. (\ref{dglap2}) can be written as
\begin{multline}
T_a(Q,k_t)\frac{\partial f_a(x,k_t^2)}{\partial \ln k_t^2}= \\
T_a(Q,k_t)\sum_{a'}\left[\int_x^{1-\Delta}\frac{dz}{z} P_{aa'}(z,k_t)f_{a'}\left(\frac{x}{z},k_t\right)-f_a(x,k_t^2)\int_0^{1-\Delta}zP_{a'a}(z,k_t)dz \right],
\end{multline}
which is the ``DGLAP equation" multiplied by $T_a$. The WMR uPDFs are defined as 
\begin{equation}
F_a(x,k_t^2,Q^2)=\frac{1}{k_t^2}f_a(x,k_t^2,Q^2)=\frac{1}{k_t^2}\frac{\partial}{\partial \ln k_t^2}\left[T_a(Q,k_t)f_a(x,k_t)\right].\label{def1}
\end{equation}
Collinear and unintegrated PDFs can be distinguished by the number of their arguments. In the following, uPDFs will refer indistinctly to $F_a(x,k_t^2,Q^2)$ or $f_a(x,k_t^2,Q^2)$. Integrating $F_a(x,kt^2,Q^2)$ over $k_t^2$ gives
\begin{equation}
\int_{Q_0^2}^{Q^2}dk_t^2F_a(x,k_t^2,Q^2)=f_a(x,Q^2)-T_a(Q^2,Q_0^2)f_a(x,Q_0^2),
\end{equation}
which, for $Q^2 \gg Q_0^2$, is numerically close to Eq. (\ref{inttmd}). Using Eq. (\ref{dglap2}), the WMR unintegrated PDFs can also be defined by
\begin{equation}
f_a(x,k_t^2,Q)=T_a(Q,k_t)\sum_{a'}\int_x^{1-\Delta}\frac{dz}{z}P_{aa'}(z,k_t)f_{a'}\left(\frac{x}{z},k_t\right).\label{def2}
\end{equation}
The main concern of Ref. \cite{BGS} was the fact that the definitions (\ref{def1}) and (\ref{def2}) do not give the same numerical result.

\section{Discussion of the KMR/WMR uPDF definitions}\label{discc}
As explained in Ref. \cite{BGS}, two cutoff are usually used: the strong ordering (SO) cutoff
\begin{equation}
\Delta=\frac{k_t}{Q}
\end{equation}
and the  angular ordering (AO) cutoff
\begin{equation}
\Delta=\frac{k_t}{k_t+Q}.
\end{equation}
By using a cutoff dependent parton density [$D_a(x,\mu^2,\Delta)$] instead of the usual one, the authors have shown that Eqs. (\ref{def1}) and (\ref{def2}) give the same numerical result. This implies that the unintegrated PDFs also depend on the cutoff, $D_a(x,k_t^2,\mu^2,\Delta)$. However, this is not really satisfactory since we started with Eq. (\ref{inttmd}). Moreover, it is not clear how this new object should be used in practice, in the phenomenology.\\

In fact, the reason why the two definitions give different results is because Eq. (\ref{dglap2}) is not always true. Let us consider the case of the AO cutoff. In this case, $k_t>Q$ is not forbidden and the Sudakov factor can be larger than 1. In order to avoid this situation, the authors defined
\begin{equation}
T_a(Q,k_t)=1, \;\;\; k_t>Q. \label{tsup1}
\end{equation}
This equation can be written as
\begin{equation}
\widetilde{T}_a(Q,k_t)=\Theta(Q^2-k_t^2)T_a(Q,k_t)+\Theta(k_t^2-Q^2), \label{suda2}
\end{equation}
where $\Theta$ is the Heaviside function. In the previous section, we mentioned that one has to be careful with the $k_t$ dependence of the Sudakov factor. With the new Sudakov factor, the lhs. of Eq. (\ref{dglap2}) gives
\begin{multline}
\frac{\partial}{\partial \ln k_t^2}\left[\widetilde{T}_a(Q,k_t)f_a(x,k_t)\right]=\left[k_t^2T_a(Q,k_t)\frac{\partial}{\partial k_t^2}\Theta(Q^2-k_t^2)+\right.\\
\left.+\Theta(Q^2-k_t^2)\frac{\partial}{\partial \ln k_t^2}T_a(Q,k_t)+k_t^2\frac{\partial}{\partial k_t^2}\Theta(k_t^2-Q^2)\right]f_a(x,k_t)+\widetilde{T}_a(Q,k_t)\frac{\partial}{\partial \ln k_t^2}f_a(x,k_t).
\end{multline}
Having in mind that $\left< \frac{d}{dx}\Theta(x-y),\phi\right> = -\left< \frac{d}{dx}\Theta(y-x),\phi\right>=\left<\delta(x-y),\phi\right>$ and that $T_a(Q,Q)=1$, we see that the first and third terms in the bracket will cancel. Taking the derivative of the second term in the bracket and rewriting it in terms of $\widetilde{T}_a$, we have
\begin{multline}
\frac{\partial}{\partial \ln k_t^2}\left[\widetilde{T}_a(Q,k_t)f_a(x,k_t)\right]=\widetilde{T}_a(Q,k_t)f_{a}(x,k_t)\sum_{a'}\int_0^{1-\Delta}dz\,zP_{a'a}(z,k_t)\\
-\Theta(k_t^2-Q^2)f_{a}(x,k_t)\sum_{a'}\int_0^{1-\Delta}dz\,zP_{a'a}(z,k_t)+\widetilde{T}_a(Q,k_t)\frac{\partial}{\partial \ln k_t^2}f_a(x,k_t).
\end{multline}
Finally, using the DGLAP equation for the last term, we get
\begin{multline}
\frac{\partial}{\partial \ln k_t^2}\left[\widetilde{T}_a(Q,k_t)f_a(x,k_t)\right]=\widetilde{T}_a(Q,k_t)\sum_{a'}\int_x^{1-\Delta}\frac{dz}{z}P_{aa'}(z,k_t)f_{a'}\left(\frac{x}{z},k_t\right) \\
-\Theta(k_t^2-Q^2)f_{a}(x,k_t)\sum_{a'}\int_0^{1-\Delta}dz\,zP_{a'a}(z,k_t), \label{newrel}
\end{multline}
showing that the definitions (\ref{def1}) and (\ref{def2}) (with $T_a$ replaced by $\widetilde{T}_a$) are not equivalent. There is then no need for these definitions to give the same numerical result, and no need for the cutoff-dependent distribution functions.\footnote{This does not mean that this object is devoid of interest. In any case, a cutoff will appear in the numerical implementation of unintegrated PDFs based on Eq. (\ref{inttmd}).}

\section{The $k_t$ dependence of WMR uPDFs}\label{sectrue}
In this section we want to insist on the conclusion reached in Ref. \cite{gui2}, that is, that the main contribution to the $p_t$ distribution of one heavy flavor is given by $Qg\rightarrow Qg$, not $gg\rightarrow Q\bar{Q}$ (for variable-flavor-number schemes). Using the KMR/WMR parametrization and the AO cutoff, one gets a satisfying result with  $gg\rightarrow Q\bar{Q}$ alone, because of the too large $k_t$ tail of the distribution. Of course, there is no reason to stop the calculation at this point, and the $Qg\rightarrow Qg$ contribution should also be computed. Doing this, the cross section for heavy-quark production will completely overshoot the data (or NLO calculations \cite{fonll} for a bare heavy quark), as we will demonstrate below.\\

In the opposite case, artificially cutting the WMR uPDFs at $k_t>Q$ and adding up the $Qg$ and $gg$ contributions gives an excellent result (see  figure 11 of Ref. \cite{gui2}). The present work has been motivated by the fact that the $k_t$ distribution of the WMR uPDFs presented in Ref. \cite{BGS} (for the SO cutoff; figure 1, left, red curve) is very similar to the cut-WMR uPDFs used in Ref. \cite{gui2}. This implies that, using the SO cutoff, the $gg\rightarrow Q\bar{Q}$ contribution will not be sufficient, and taking into account $Qg\rightarrow Qg$ will be necessary to bring agreement with data, as it should be. Leaving this discussion for later, we continue with the analysis of Eq. (\ref{newrel}) and of the AO cutoff.\\

\begin{figure}[!h]
\centering
\includegraphics[width=10.0cm]{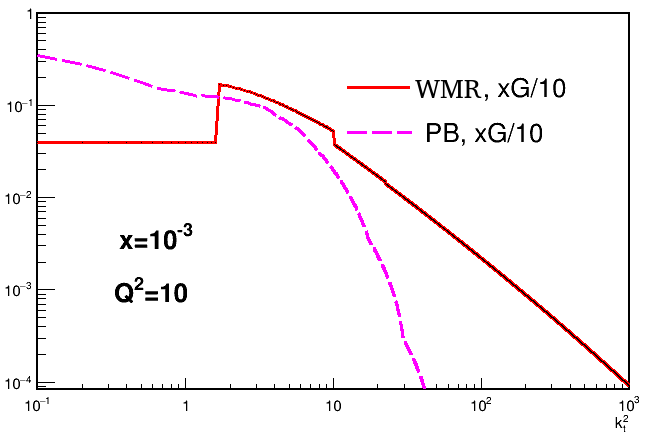}
\caption{WMR unintegrated gluon density as a function of $k_t^2$, showing a discontinuity at $k_t^2= Q^2$. Here $G=f_g/k_t^2$, with $f_g$ given by Eq. (\ref{def3}). It is compared to the PB uPDFs \cite{pbnlo}, which give an accurate result for the heavy-quark $p_t$ distribution \cite{gui2}. \label{kmrnn}}
\end{figure}

We first note that $\widetilde{T}_a(Q,Q)=1$. Then, integrating the lhs. of Eq. (\ref{newrel}) gives a result that is numerically close to Eq. (\ref{inttmd}). Consequently, a possible correct definition of the WMR uPDFs is
\begin{align}
f_a(x,k_t^2,Q)&=\widetilde{T}_a(Q,k_t)\sum_{a'}\int_x^{1-\Delta}\frac{dz}{z}P_{aa'}(z,k_t)f_{a'}\left(\frac{x}{z},k_t\right) \nonumber\\
&-\Theta(k_t^2-Q^2)f_{a}(x,k_t)\sum_{a'}\int_0^{1-\Delta}dz\,zP_{a'a}(z,k_t).\label{def3}
\end{align} 
This distribution is displayed in Fig. \ref{kmrnn}, for $x=10^{-3}$ and $Q^2=10$ GeV$^2$. 
Compared to Eq. (\ref{def2}), it receives a negative contribution at $k_t>Q$. Then, it presents a discontinuity at $k_t=Q$, identical to the result shown in Ref. \cite{BGS} (the dashed blue line in the right panel of Fig. 1), obtained from the definition (\ref{def1}). This shows the equivalence of Eqs. (\ref{def1}) and (\ref{def3}), without the need for a cutoff-dependent parton density; the issue was that Eq. (\ref{dglap2}) is incorrect for the Sudakov factor defined in Eq. (\ref{suda2}).\\

\begin{figure}[!h]
\centering
\includegraphics[width=10.0cm]{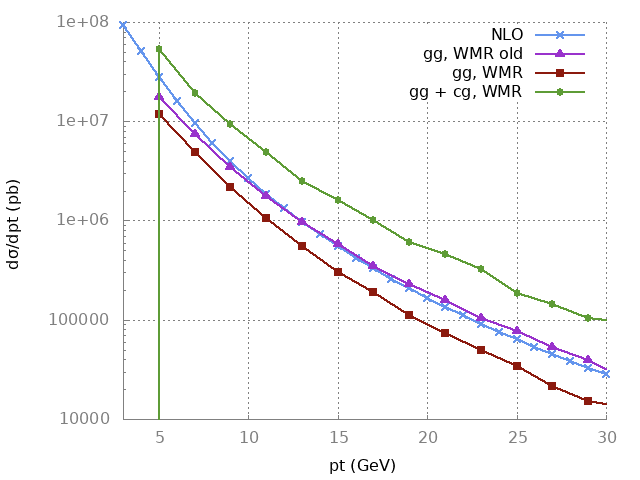}
\caption{NLO calculations \cite{fonll} for the charm $p_t$ distribution, compared to results obtained with \textsc{KaTie} \cite{katie} and the WMR uPDFs. ``WMR old" refers to Eq. (\ref{def2}) [or equivalently to Eq. (\ref{def4})], while ``WMR" is for Eq. (\ref{def3}). \label{kmrpt}}
\end{figure}

The main theoretical issue with the AO cutoff is that there is an infinite number of nonequivalent definitions of the uPDFs which do agree with Eq. (\ref{inttmd}). Indeed, we can always add $\Theta(k_t^2-Q^2)A(x,k_t^2,Q^2)$ to the definition (\ref{def3}), where $A(x,k_t^2,Q^2)$ is any function.\footnote{This is due to the fact that in Eq. (\ref{inttmd}), the uPDFs are only integrated up to $Q^2$. In the parton model, as defined in \cite{col}, the relation is $f(\xi)=\int_0^{\infty}d^2k_tf(\xi,k_t^2)$.} In particular, another correct definition is
\begin{align}
f_a(x,k_t^2,Q)&=\widetilde{T}_a(Q,k_t)\sum_{a'}\int_x^{1-\Delta}\frac{dz}{z}P_{aa'}(z,k_t)f_{a'}\left(\frac{x}{z},k_t\right) \nonumber\\
&=\frac{\partial}{\partial \ln k_t^2}\left[\widetilde{T}_a(Q,k_t)f_a(x,k_t)\right]+\Theta(k_t^2-Q^2)f_{a}(x,k_t)\nonumber\\
&\times \sum_{a'}\int_0^{1-\Delta}dz\,zP_{a'a}(z,k_t) \label{def4}
\end{align}
These definitions differ for $k_t>Q$, and lead to significant differences for the heavy-quark cross section, as shown in Fig. \ref{kmrpt}. The consequence is a loss of predictability for observables sensitive to the region $k_t>Q$. Note that we can also choose the function $A(x,k_t^2,Q^2)$ such that $f_a(x,k_t^2,Q)=0$ for $k_t>Q$. It is clear that, with the AO cutoff, Eq. (\ref{inttmd}) is not enough to fix the definition of the KMR/WMR uPDFs. An extra condition could be that we want the distribution and its first derivative to be continuous at large $k_t$\footnote{In any case, the distribution has a discontinuity at small $k_t$.}. This corresponds to the definition given in Eq. (\ref{def4}). A better condition is that numerical calculations should be in agreement with data once all contributions have been taken into account at a given order.\\

However, these two conditions are not compatible. The distributions obtained from Eq. (\ref{def4}) or Eq. (\ref{def3}) are too large for $k_t>Q$. The contribution $gg\rightarrow Q\bar{Q} \; +\; Qg\rightarrow Qg$  overestimates the NLO calculations \cite{fonll} for the heavy-quark $p_t$ distribution, as shown in Fig. \ref{kmrpt}.
These results have been obtained with the \textsc{KaTie} event generator \cite{katie}, designed for $k_t$-factorization calculations with off-shell matrix elements. The setup is identical to the one described in Ref. \cite{gui2}. In particular, we use the conventional factorization scale $\mu=(p_t^c+p_t^X)/2$, with $c$ referring to the outgoing charm and $X$ to the other particle. The charm mass has been set to 0 in the process $cg\rightarrow cg$. ``WMR old" and ``WMR" refer to the definitions (\ref{def2}) [or equivalently to Eq. (\ref{def4})] and (\ref{def3}), respectively. As expected, the latter gives a smaller $gg$ contribution due to the smaller unintegrated gluon density at $k_t>Q$. However, we can see that the $gg +cg$ contribution still overestimates NLO calculations. In fact, the $cg\rightarrow cg$ contribution alone already overshoots the NLO result, showing that the overestimation is not due to a double counting between $gg\rightarrow c\bar{c}$ and $cg\rightarrow cg$; rather, it is a consequence of the fact that the AO uPDFs are ill defined. In Ref. \cite{gui2}, it has been shown that the same calculations done with the PB uPDFs \cite{pbnlo} do a good job.\\

We now discuss the KMR/WMR prescription with the SO cutoff, and we will see that it solves all of these issues. In this case, the condition $x<1-\Delta$ implies that
\begin{equation}
k_t\leq Q(1-x) \leq Q, \label{condkt}
\end{equation}
giving a Sudakov factor smaller than 1. The condition $x<1-\Delta$ is true regardless of the uPDF definition, and it can be maintained explicitly by a factor of $\Theta(Q^2-k_t^2)$ in Eqs. (\ref{def1}) and (\ref{def2}). In this case, Eq. (\ref{dglap2}) is true and both definitions give the same result, namely, a distribution with a sharp cutoff for $k_t>Q$. Consequently, the SO cutoff eliminates the issue of the multiple uPDF  definitions.

\begin{figure}[!h]
\centering
\includegraphics[width=10.0cm]{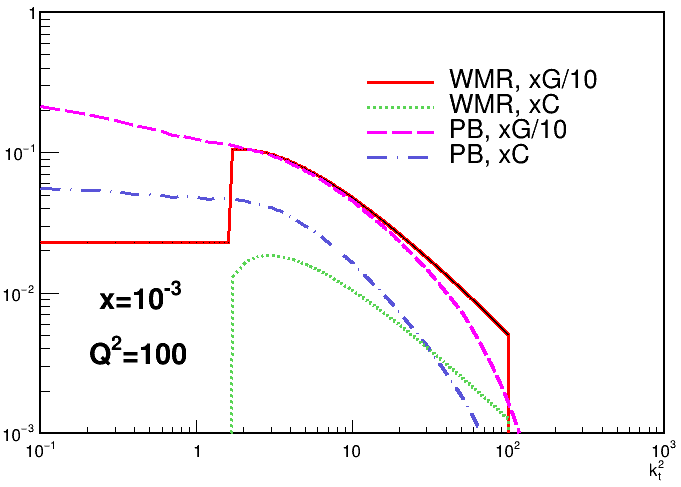}
\caption{Charm and gluon uPDFs obtained with the WMR prescription and the SO cutoff, compared to the PB uPDFs. \label{SOkmr}}
\end{figure}

\begin{figure}[!h]
\centering
\includegraphics[width=10.0cm]{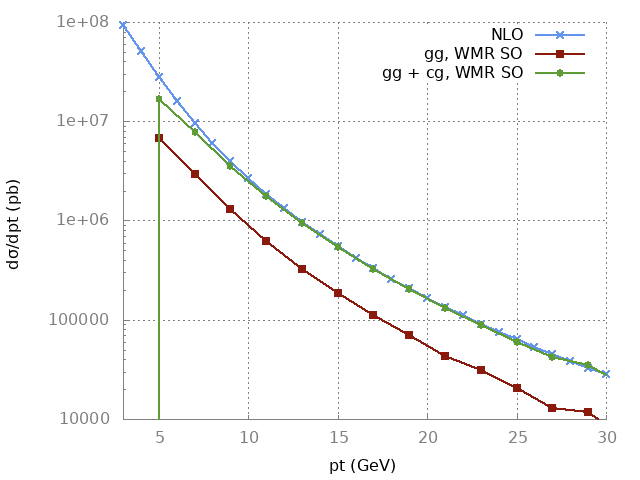}
\caption{Charm distribution obtained with \textsc{KaTie} and the WMR uPDFs presented in Fig. \ref{SOkmr}. \label{ptkmrso}}
\end{figure}

In Fig. \ref{SOkmr}, we show the $k_t$ dependence of the WMR uPDFs computed with this cutoff.
For $k_t>1$ GeV, these distributions are quite similar to the PB uPDFs, and we can anticipate that they will give similar results. This is indeed the case, as shown in Fig. \ref{ptkmrso}.
As expected, the $gg$ contribution undershoots the NLO calculations for the charm $p_t$ distribution. It is only after including the $cg$ contribution (the main one) that we obtain agreement between them. Note that we still have to include the $q\bar{q}\rightarrow Q\bar{Q}$ and $cq\rightarrow cq$ processes, which are negligible and small, respectively \cite{gui2} (at least in this kinematical range).\\

Note the small difference between the slope of the $gg + cg$ contribution (Fig. \ref{ptkmrso}, green line) and the slope of the $gg$ contribution (Fig. \ref{kmrpt}, purple line) obtained with the AO cutoff. The former is harder and exactly follows NLO calculations. However, this small difference should not be overinterpreted. As explained before, we have neglected small contributions and the full calculation could exhibit a slightly modified slope. Moreover, the slope also depends on the choice made for the factorization scale.

\section{Remarks and discussions}

\subsection{Double counting}
The result obtained with the AO KMR uPDFs, which shows an overestimation of the heavy-quark production, as illustrated in Fig. \ref{kmrpt}, may look suspicious. In this section we demonstrate that this overestimation is not due to a double counting. As discussed earlier, it is simply a consequence of the fact that the AO KMR uPDFs are ill defined. \\

\begin{figure}[!h]
\centering
\includegraphics[width=3.0cm]{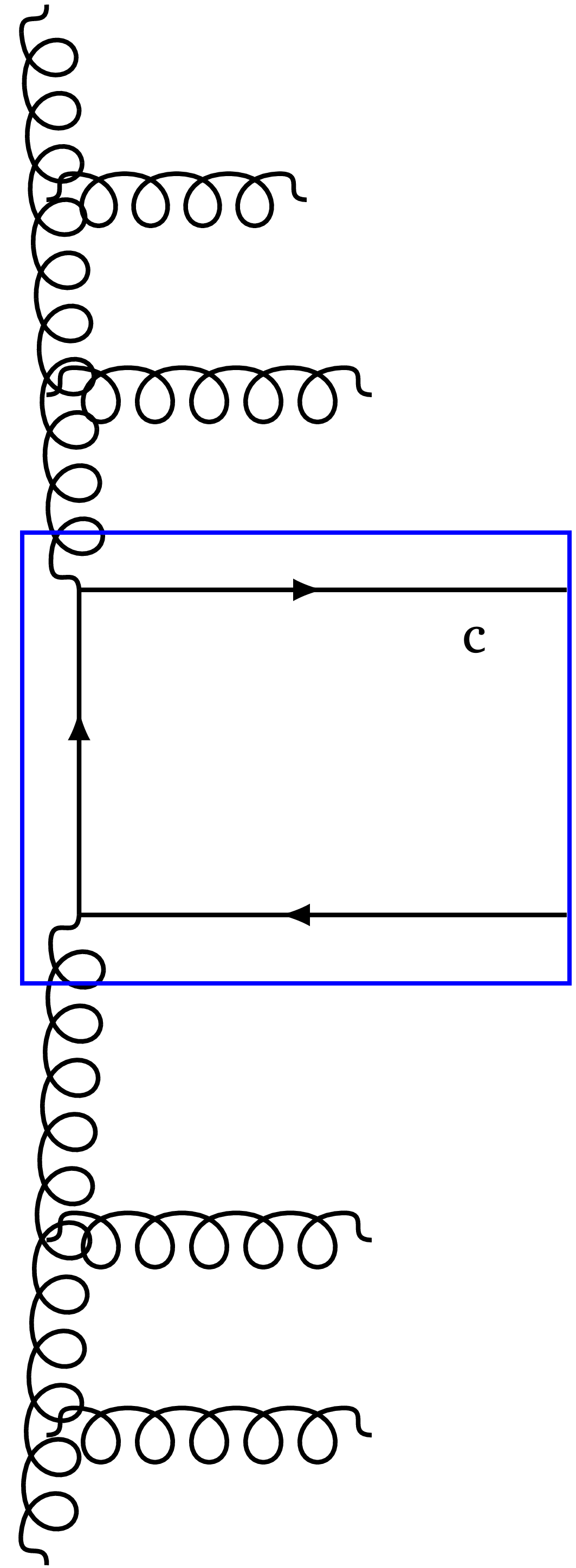}
\includegraphics[width=3.0cm]{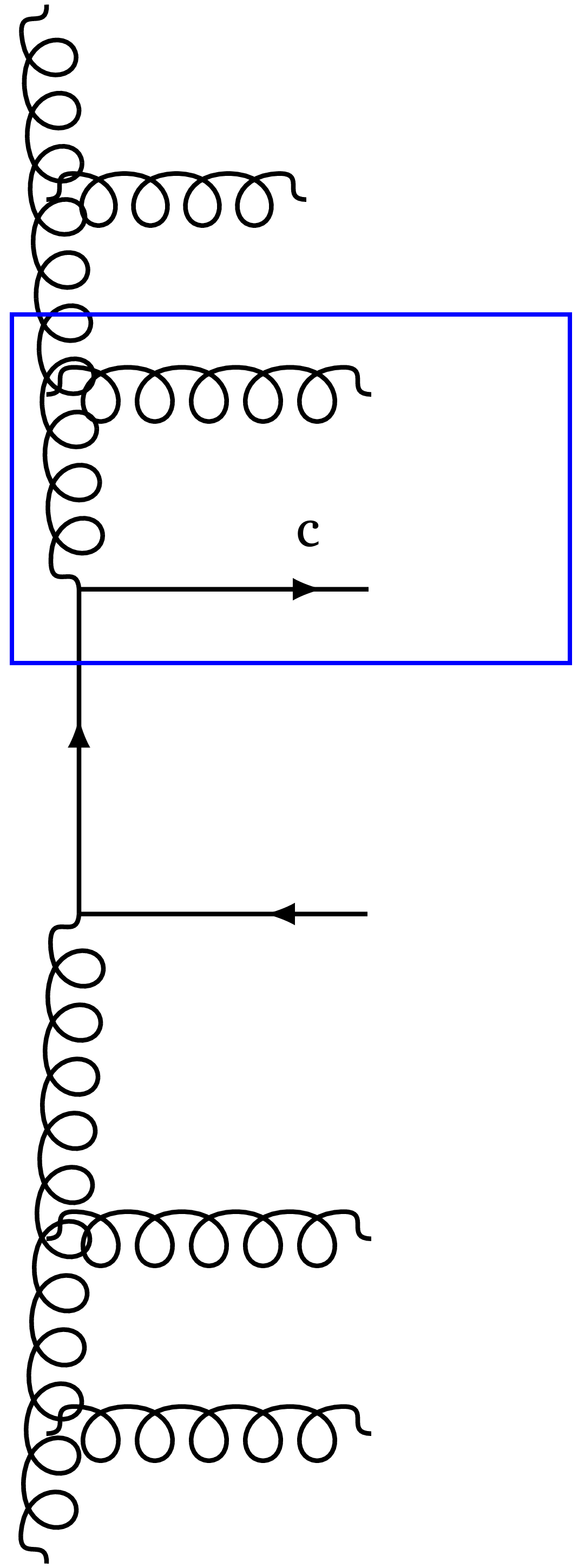}
\caption{Example of Feynman diagrams for $gg\rightarrow c\bar{c}$ (left) and $cg \rightarrow cg$ (right).  The blue squares indicate the $2\rightarrow 2$ interactions. \label{dcdia}}
\end{figure}

\begin{figure}[!h]
\centering
\includegraphics[width=2.5cm]{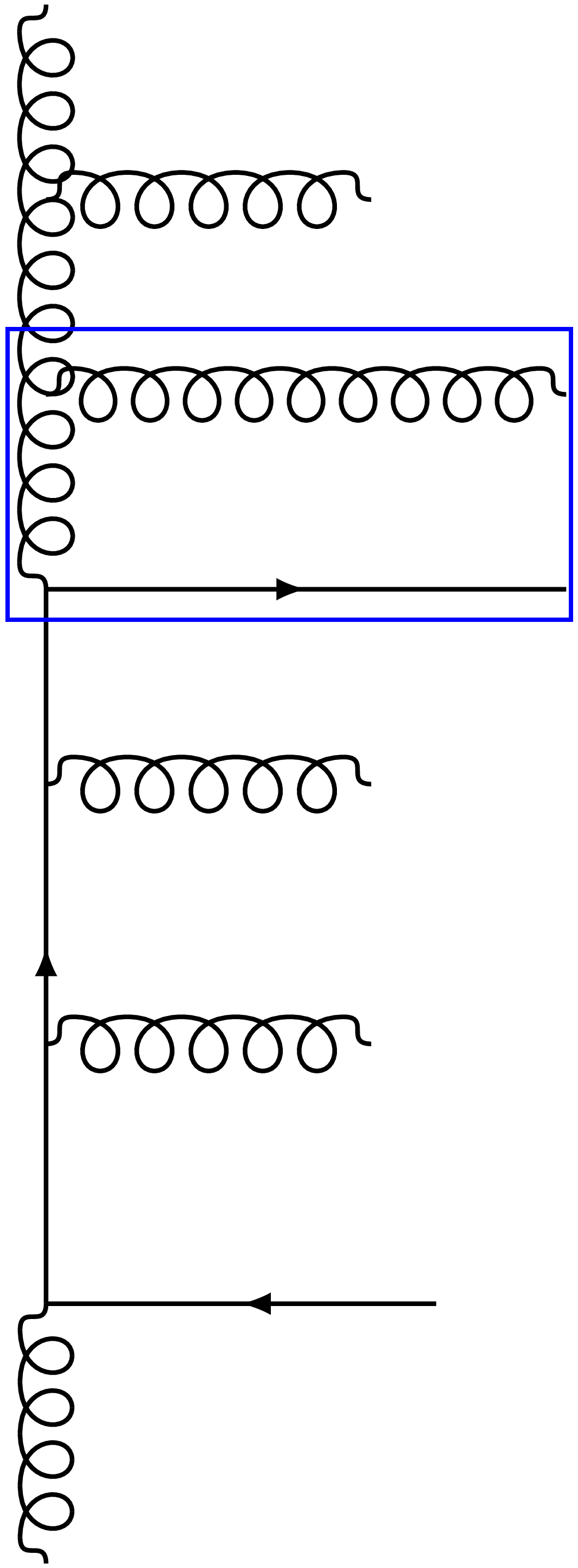}
\caption{Example of a Feynman diagram for $cg \rightarrow cg$ showing no similarity with the $gg\rightarrow c\bar{c}$ diagrams. \label{dddia}}
\end{figure}

One might suspect a double counting because of the similarity between the diagrams shown in Fig. \ref{dcdia}. The blue squares indicate the $2\rightarrow 2$ matrix elements for $gg\rightarrow c\bar{c}$ (left) and $cg \rightarrow cg$ (right).
If one forgets about the blue squares, these diagrams look exactly the same. However, there are not the same because each blue square contains in fact three Feynman diagrams (for the $s$, $t$, and $u$ channels), and because the phase space is not the same (but a partial overlapping could be possible). Even though they look similar, they correspond to two different physical processes: the collision of two gluons and the collision of a gluon with a charm quark. Note also that the $cg$ process includes an infinite number of Feynman diagrams that are not similar to the one for the leading-order $gg\rightarrow c\bar{c}$ process; an example is shown in Fig. \ref{dddia}.\\

A clear and simple argument showing that the overestimation is not due to a double counting between the diagrams shown in Fig. \ref{dcdia} is the observation that with the AO KMR uPDFs, the $cg$ process alone already overestimates the charm cross section.\\

\subsection{$2\rightarrow 1$ vs. $2\rightarrow 2$ matrix elements}
\begin{figure}[!h]
\centering
\includegraphics[width=3.0cm]{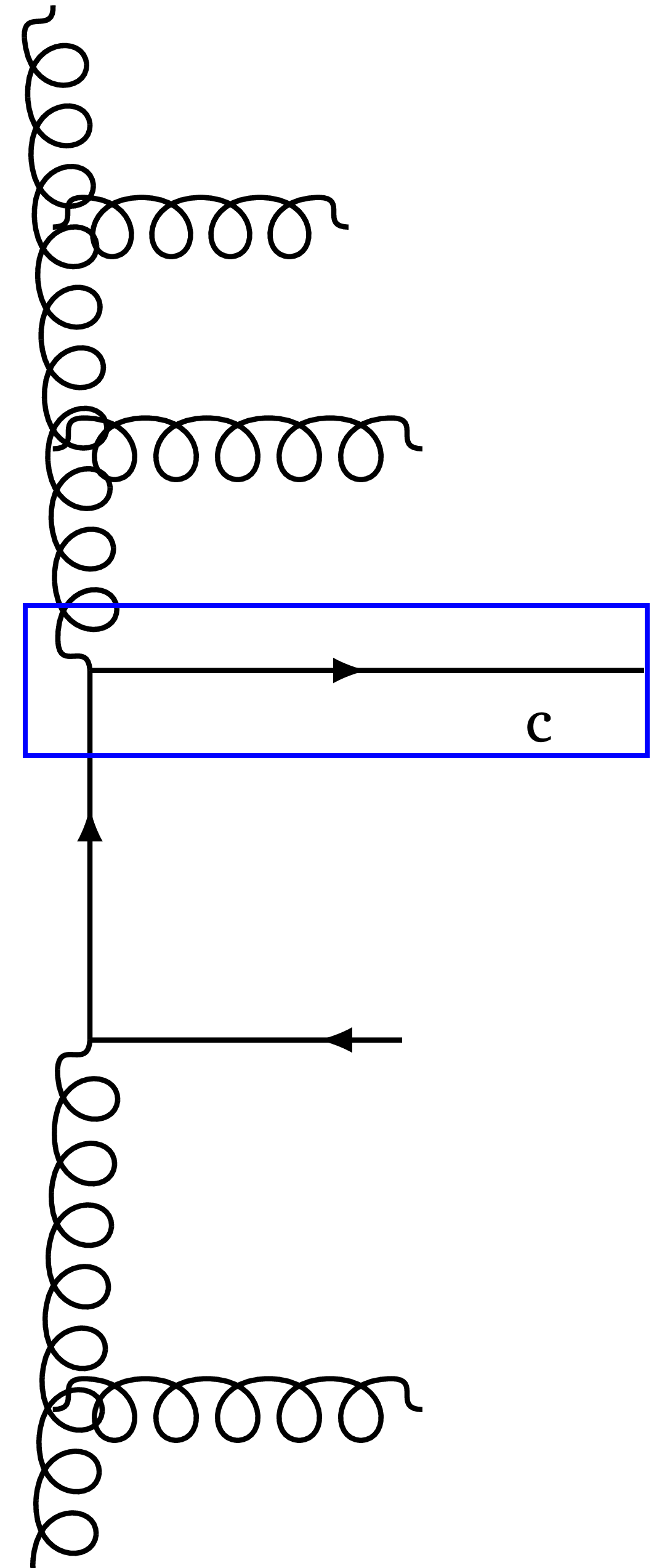}
\caption{Example of Feynman diagrams for the off-shell matrix element $gc\rightarrow c$. \label{dia21}}
\end{figure}

In $k_t$ factorization, a $2\rightarrow 1$ process is kinematically allowed, and calculations for D meson production were performed in this way 10 years ago \cite{KnShSa}. A Feynman diagram for this process is shown in Fig. \ref{dia21}.
Even if the $cg\rightarrow c$ process seems to be the true leading order for D meson production, it is in fact approximately equivalent to the $2\rightarrow 2$ process. This can be seen by considering one of the outgoing partons of the $2\rightarrow 2$ process as being part of the evolution (compare for instance Fig. \ref{dia21} with Fig. \ref{dcdia}, left). However, these two formalisms are not completely equivalent, and calculations performed in Ref. \cite{KnShSa} did not include s-channel Feynman diagrams (at high energies, these diagrams are indeed negligible). 

We believe that recent calculations using the $2\rightarrow 2$ processes give a better result, but it would be interesting to perform a precise comparison between these two points of view\footnote{A precise comparison would involve the use of the same uPDFs set. Note that in Ref.  \cite{KnShSa}, uPDFs of the KMR type were used.}.

\subsection{On the $k_t$ factorization}\label{secrem}
We have mentioned that the KMR/WMR uPDFs used with the AO cutoff are ill defined. This issue is not restricted to the KMR/WMR prescriptions, and any uPDFs with such large $k_t$ tail at $k_t>Q$ will encounter the same problem. We believe that this issue could be related to the fact that there is no proof of the $k_t$ factorization. The consequence is the absence of a precise definition for the unintegrated parton densities. In the $k_t$-factorization formalism, these functions should only approximately respect the relation (\ref{inttmd}). If, for instance, one instead chooses the relation to be
\begin{equation}
f(x,Q^2)=\int_0^{\infty}dk_t^2 F(x,k_t^2,Q^2),
\end{equation}
the issue of ill-defined uPDFs would disappear, because the large $k_t$ tail of the distribution would be constrained. The Blümlein's uPDFs obey this relation; see, for instance, Ref. \cite{AnBaBa}. 

Note also that the PB and WMR uPDFs were recently compared in Ref. \cite{HaKeLe}. The authors have shown that in the case of the PB uPDFs, integrating up to $Q^2$ or up to infinity gives a result compatible with collinear PDFs in both cases (see Fig. 6 in Ref. \cite{HaKeLe}). In the opposite case, the WMR uPDFs used with the angular-ordering cutoff give a numerical result compatible with the collinear PDFs only if the integration is stopped at $Q^2$. Otherwise, the numerical result overestimates the collinear PDFs, in particular at small x, showing the non-negligible role played by the tail of the distribution for $k_t>Q$.

\section{Conclusion}
In this paper we discussed the KMR and WMR prescriptions for uPDFs, and we underlined the fact that several recent studies that used the ``KMR" prescription in fact used the WMR one. We have seen that only the WMR prescription gives the correct DGLAP equation. 

Then, we addressed the issue of the apparently mathematically equivalent uPDF definitions giving different numerical results, mentioned in Ref.  \cite{BGS}. We have demonstrated that, with the Sudakov factor used in Ref. \cite{BGS}, these definitions were in fact not equivalent, and we gave the correct relation, Eq. (\ref{newrel}).

We have seen that the WMR prescription leads to significant differences for the charm cross sections, depending on the choice made for the cutoff. With the AO cutoff, the contribution $gg + cg$ completely overshoots the NLO calculations (Fig. \ref{kmrpt}) and the uPDFs are not uniquely defined by Eq. (\ref{inttmd}), leading to a loss of predictability. However, the issue arising from the region $k_t>\mu$, where $\mu$ is the factorization scale, observables obtained after a $k_t$ integration in the region $k_t \in [0,E]$, with $E \sim \mu$, are safe. This is the case, for instance, for Drell-Yan production at large $\hat{s}$. In section \ref{secrem}, we discussed the potential relation between this issue and the lack of a proper definition for unintegrated PDFs. 

In the opposite case, the SO cutoff avoids these issues. It gives satisfying numerical results (Fig. \ref{ptkmrso}), in agreement with those obtained in Ref. \cite{gui2}. In particular, using the (SO) WMR uPDFs, we confirmed that the main contribution to heavy-quark production is given by $Qg\rightarrow Qg$, the $gg$ contribution alone being a factor of $\sim 3$ below NLO calculations. Compared to the AO cutoff, the obtained  $k_t$ distributions are closer to other uPDFs sets, e.g., the PB uPDFs.

Unfortunately, the majority of phenomenological papers use the AO cutoff. Calculations are done including only the $gg$ contribution [with the gluon unintegrated density built from Eq. (\ref{def4})], giving an (accidental) reasonable agreement with data. The fact that the other contributions are not including is not even mentioned. One of the unpleasant consequences is to convince the reader that the main contribution to heavy-quark production is the $gg$ contribution. Then, using another correct uPDF set, e.g., the PB one \cite{pbnlo}, and including only the $gg$ contribution, leads to the erroneous conclusion that this set does not work. This was the case, for instance, in Ref. \cite{MacSzc}, where the PB and KMR uPDFs were discussed. In that paper, it is said that ``a new Parton-Branching (PB) uPDF strongly underestimates the same experimental data". However, it was shown in Ref. \cite{gui2} that, once all contributions have been added up, the PB uPDFs give in fact a good description of the heavy-quark $p_t$ distribution.

\section*{Acknowledgments}
We would like to thank T. Mineeva for valuable comments. We acknowledge support from Chilean FONDECYT Iniciaci\'on grant 11181126. We acknowledge support by the Basal project FB0821.

%\section*{References}


\begin{thebibliography}{9}
\bibitem{BGS}K. Golec-Biernat and A. M. Sta\'sto, ``On the use of the KMR unintegrated parton distribution functions", Phys. Lett. B 781 (2018) 633-638.
\bibitem{JiMaYu1} X.-d. Ji, J.-P. Ma, and F. Yuan, ``QCD factorization for spin-dependent cross sections in DIS and Drell-Yan processes at low transverse momentum", Phys. Lett. B597, 299 (2004).
\bibitem{JiMaYu2} X.-d. Ji, J.-p. Ma, and F. Yuan, ``QCD factorization for semi-inclusive deep-inelastic scattering at low transverse momentum", Phys. Rev. D71, 034005 (2005).
\bibitem{BaBoDi} A. Bacchetta, D. Boer, M. Diehl, and P. J. Mulders, ``Matches and mismatches in the descriptions of semi-inclusive processes at low and high transverse momentum", JHEP 0808, 023 (2008).
\bibitem{col} John Collins, ``Foundations of perturbative QCD", Camb. Monogr. Part. Phys. Nucl. Phys. Cosmol. 32 (2011) 1-624.
\bibitem{ktFac} J.C. Collins and R.K. Ellis, ``Heavy-quark production in very high energy hadron collision'', Nucl. Phys. B 360 (1991) 3-30.
\bibitem{ktFac2} S. Catani, M. Ciafaloni and F. Hautmann, ``High energy factorization and small-$x$ heavy flavour production'', Nucl. Phys. B 366(1991) 135-188.
\bibitem{semihard1} L. Gribov, E. Levin, M. Ryskin, ``Semihard processes in QCD'', Phys. Rep. 100 (1983) 1.
\bibitem{semihard2} E. M. Levin, M. G. Ryskin, Y. M. Shabelski, A. G. Shuvaev, ``Heavy quark production in semihard nucleon interaction'', Sov. J. Nucl. Phys. 53 (1991) 657.
\bibitem{KMR1}M. A. Kimber, A. D. Martin, M. G. Ryskin, ``Unintegrated parton distributions'', Phys. Rev. D 63 (2001) 114027.
\bibitem{KMR2} G. Watt, A.D. Martin and M.G. Ryskin, ``Unintegrated parton distributions and inclusive jet production at HERA'', Eur. Phys. J. C 31 (2003) 73.
\bibitem{gui2} B. Guiot, ``Heavy-quark production with $k_t$-factorization: The importance of the sea-quark distribution", Phys.Rev. D99 (2019) no.7, 074006.
\bibitem{ElStWe} R. K. Ellis, W. J. Stirling, and B. R. Webber, ``QCD and Collider Physics", Cambridge university press (1996).
\bibitem{fonll}   M.~Cacciari, M.~Greco and P.~Nason,``The p(T) spectrum in heavy-flavor hadroproduction", JHEP {\bf 05} (1998) 007;\\
                  M.~Cacciari, S.~Frixione and P.~Nason, ``The p(T) spectrum in heavy-flavor photoproduction”, JHEP {\bf 03} (2001) 006.
\bibitem{pbnlo} A. Bermudez Martinez, P. Connor, F. Hautmann, H. Jung, A. Lelek, V. Radescu, and R. Zlebcik, ``Collinear and TMD parton densities from fits to precision DIS measurements in the parton branching method'',  Phys. Rev. D 99, 074008 (2019).
\bibitem{katie} A. Van Hameren, ``\textsc{KaTie} : For parton-level event generation with $k_T$-dependent initial states'', Comput.Phys.Commun. 224 (2018) 371-380.
\bibitem{KnShSa} B.A. Kniehl, A.V. Shipilova, V.A. Saleev, ``Open charm production at high energies and the quark Reggeization hypothesis", Phys. Rev. D 79, 034007.
\bibitem{AnBaBa} Bo Andersson et al., ``Small x Phenomenology. Summary and Status", Eur.Phys.J. C25 (2002) 77-101.
\bibitem{HaKeLe} F. Hautmann, L. Keersmaekers, A. Lelek and A. M. van Kampen, ``Dynamical resolution scale in transverse momentum distributions at the LHC", Nucl. Phys. B949, 114795 (2019).
\bibitem{MacSzc} R. Maciula and A. Szczurek, ``Consistent treatment of charm production in higher-orders at tree-level within $k_T$-factorization approach", Phys. Rev. D 100, 054001 (2019).
\end{thebibliography}
\end{document}